\documentclass[12pt]{article}
\usepackage{amsmath}
\usepackage{graphicx}
\begin{document}
\baselineskip 18pt

%%%%%%%%%%%%%%%%%%%%%%%%%%%%%%%%%%%%%%%%%%%%%%%%%%

\begin{titlepage}
\begin{flushright}
UT-07-17
\end{flushright}
\begin{center}
{\Large \textbf{
  Testing the seesaw mechanism at collider energies \medskip \\
  in the Randall-Sundrum model
}}
\vskip 20pt
Hiroto Nakajima
\vskip 20pt
\textit{Department of Physics, University of Tokyo, \\ Tokyo 113-0033, Japan}
\vskip 20pt
\abstract{
  The Randall-Sundrum model with gauge fields and fermions in the bulk
  has several attractive features
  including the gauge coupling unification,
  a candidate for the dark matter,
  and an explanation for the hierarchical Yukawa couplings.

  In this paper, we point out that the 1st Kaluza-Klein modes
  of right-handed neutrinos may be produced at colliders,
  for both Dirac and Majorana neutrino cases.
  Furthermore, we can see whether the neutrino masses
  are Dirac type or Majorana type
  from the mass spectrum of KK particles.
}
\end{center}
\end{titlepage}

%%%%%%%%%%%%%%%%%%%%%%%%%%%%%%%%%%%%%%%%%%%%%%%%%%

\setcounter{page}{2}
\section{Introduction}

Recently, models with extra dimensions were suggested
as solutions to the gauge hierarchy problem.
Amongst these models, the Randall-Sundrum (RS) model
\cite{Randall:1999ee} attracted particular attention.
In the RS model, the hierarchy between the electroweak and the Planck scales
is explained by a warped extra dimension.
Originally, all Standard Model (SM) fields are localized on the TeV brane.
However, it was realized that it is sufficient to localize only the Higgs
on the TeV brane to solve the hierarchy problem.
Moreover, placing gauge fields and fermions in the bulk
brings several attractive features:
unification of the gauge couplings at high scale
\cite{Pomarol:2000hp}$-$\cite{Agashe:2002pr},
a candidate for the dark matter
\cite{Agashe:2004ci, Agashe:2004bm},
and a new interpretation for the hierarchical Yukawa couplings
\cite{Gherghetta:2000qt, Dooling:2000ky}.

The hierarchical Yukawa couplings are explained
by the overlaps of the Higgs and fermions.
Recently, certain types of configurations of fermions are obtained
\cite{Moreau:2006np}
that reproduce not only the fermions mass matricies,
but also satisfy the Flavor Changing Neutral Current (FCNC) constraints
for $m_{A}^{(1)} \ge 1$ TeV
\footnote{
  Here $m_{A}^{(1)}$ denotes the mass of the 1st KK mode of gauge boson
 (see Appendix).
}.
These models also sufficiently suppress unwanted non-renormalizable operators
except for $B$ and $L$ breaking ones.

In a recent work \cite{Nakajima:2007uk}, we showed that
we can not suppress the operators for $B$ and $L$ breaking
by the configurations of fermions that reproduce the realistic mass matricies.
These unwanted operators are easily suppressed by discrete gauge symmetries
\cite{Ibanez:1991hv},
but it is then expected that the seesaw mechanism \cite{seesaw} does not work.
However, we found that the seesaw mechanism does work
to explain the observed small mass of neutrinos,
if $L$ is broken on the Planck brane \cite{Nakajima:2007uk}.

In this paper, we point out that the 1st Kaluza-Klein (KK) modes
of right-handed neutrinos may be produced at $e^{+} e^{-}$ colliders,
for both Dirac and Majorana neutrino cases.
In 4D theories, it is difficult to produce right-handed neutrinos at colliders,
since they have large masses ($\sim 10^{15}$ GeV)
or small Yukawa couplings ($\sim 10^{-12}$).
In the RS model, KK fermions have masses of order TeV and Yukawa couplings of order 1,
which enables the production of KK right-handed neutrinos at colliders.
Furthermore, we can determine the bulk configurations of fermions
by comparing the masses of KK gauge bosons and KK fermions.
Thus we can see whether the neutrino masses are Dirac type or Majorana type
from the mass spectrum of KK particles.

%%%%%%%%%%%%%%%%%%%%%%%%%%%%%%%%%%%%%%%%%%%%%%%%%%

\section{Setup}

The metric of the RS model is
\begin{align}
  ds^{2} = e^{-2\sigma} \eta_{\mu\nu} dx^{\mu} dx^{\nu} + dy^{2}
\end{align}
where $\sigma = k|y|$,
and $k \sim M_{P} \sim 10^{18}$ GeV is the AdS curvature.
The fifth dimension $y$ is compactified on an orbifold $S^{1}/Z_{2}$.
Two 3-branes reside at the fixed points $y = 0$ and $y = \pi R$,
which are referred to as the Planck brane and the TeV brane, respectively.

We assume that the SM gauge bosons and fermions are in the bulk,
and Higgs is on the TeV brane.
The kinetic term of Higgs is
\begin{align}
    S_\mathrm{Higgs}
& = \int d^{4}x \, dy \sqrt{g_{MN}} \, g^{\mu\nu}
    D_{\mu} H(x) D_{\nu} H(x) \delta(y - \pi R) \notag \\
& = \int d^{4}x \, \eta^{\mu\nu}
    D_{\mu} \tilde{H} D_{\nu} \tilde{H},
\end{align}
where $\tilde{H} = e^{- k \pi R} H$ is a canonically normalized field.
Thus we have $\langle H \rangle = e^{k \pi R} \, v / \sqrt{2}$.

The hierarchical Yukawa couplings are explained by the configurations of fermions.
For example, Yukawa interaction terms of the charged leptons are
\begin{align}
    S_\mathrm{Yukawa}
& = \int d^{4}x \, dy \sqrt{g_{MN}} \,
   (\lambda_{e5})_{ij} H \bar{\ell}_{i}(x,y) e_{j}(x,y) \delta(y - \pi R) \notag \\
& = \int d^{4}x \, (\lambda_{e5})_{ij} \, T_{m}(c_{\ell i}) T_{n}(c_{ej}) \,
    \tilde{H} \bar{\ell}_{i}^{(m)}(x) e_{j}^{(n)}(x),
\end{align}
where $\ell_{i}^{(m)}$, $e_{j}^{(n)}$ are KK modes of leptons,
and $T_{m}(c_{\ell i})$, $T_{n}(c_{ej})$ are their couplings on the TeV brane
(see Appendix).
The lower suffixes $i$, $j$ denote generations of electroweak eigenstates,
which are often omitted in the following.
The zero modes correspond to SM leptons.
With a moderate tuning of the $O(1)$ mass parameters $c_{\ell}$ and $c_{e}$,
the hierarchical Yukawa couplings are reproduced
by effective 4D couplings $\lambda_{e5} \, T_{0}(c_{\ell}) T_{0}(c_{e})$.

We introduce a right-handed neutrino $N$ in the bulk.
The Yukawa interaction terms $H^{*} \bar{\ell} N$
generate Dirac masses for neutrinos.
In the case of Majorana neutrino,
$N$ has a Majorana mass term (which breaks $L$) on the Planck brane,
\begin{align}
    S_\mathrm{Majorana}
& = \int d^{4}x \, dy \sqrt{g_{MN}} \,
    \lambda_{5}' M(x) N^{T}(x,y) \gamma_{2} N(x,y) \delta(y) \notag \\
& = \int d^{4}x \, \lambda_{5}' \, P_{m}(c_{N}) P_{n}(c_{N}) \,
    M N^{(m)T}(x) \gamma_{2} N^{(n)}(x),
\end{align}
where $M$ is a 5D Majorana mass.
Thus we obtain 4D Lagrangians
\begin{align}
    \mathcal{L}_{KK}
& = m_{N}^{(n)}    \bar{N}^{(n)} N^{(n)}
  + m_{\ell}^{(n)} \bar{\ell}^{(n)} \ell^{(n)}
  + m_{e}^{(n)}    \bar{e}^{(n)} e^{(n)}, \\
    \mathcal{L}_\mathrm{Yukawa}
& = \lambda_{e 5}   T_{m}(c_{\ell}) T_{n}(c_{e})
    \tilde{H}     \bar{\ell}^{(m)} e^{(n)} \notag \\
& + \lambda_{\nu 5} T_{m}(c_{\ell}) T_{n}(c_{N})
    \tilde{H}^{*} \bar{\ell}^{(m)} N^{(n)}, \\
    \mathcal{L}_\mathrm{Majorana}
& = \lambda' P_{m}(c_{N}) P_{n}(c_{N})
    M N^{(m)T} \gamma_{2} N^{(n)}.
\end{align}

%%%%%%%%%%%%%%%%%%%%%%%%%%%%%%%%%%%%%%%%%%%%%%%%%%

\section{Neutrino mass}

In this section, we search for the parameters
that reproduce the heaviest left-handed neutrino mass
$\sqrt{\Delta m_\mathrm{atm}^{2}} \simeq 5 \times 10^{-2}$ eV.
We use $k = 2.4 \times 10^{18}$ GeV and $e^{k \pi R} = 2 \times 10^{15}$,
which give KK gauge boson masses $m_{A}^{(1)} \simeq 3$ TeV,
the lower bound from electroweak measurements \cite{Agashe:2003zs}.
We assume that 5D Yukawa couplings are
$\lambda_{e5}$, $\lambda_{\nu 5}$, $\lambda' \sim 1/k$,
which correspond to 4D Yukawa couplings $\lambda_{4} \sim 1$ for $c < 1/2$.
We further assume that $\lambda_{e5}$ is diagonal for simplicity.

\begin{figure}[tbp]
\begin{center}
\includegraphics[width = 0.45 \linewidth]{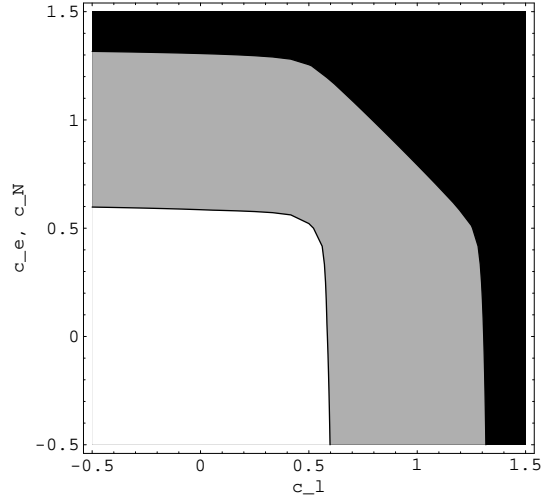}
\caption{
  The Dirac masses of zero modes.
  The contours in the figure correspond to
  $m_{\tau}$ (bottom-left) and $5 \times 10^{-2}$ eV (top-right).
  The first contour is plotted for $c_{\ell}$ and $c_{e}$,
  while the second one is for $c_{\ell}$ and $c_{N}$.
  Here we used $\lambda_{e5}$, $\lambda_{\nu 5} = 1/k$.
}
\label{Dirac}
\end{center}
\end{figure}

We have to specify the value of $c_{\ell 3}$ to evaluate the neutrino mass,
since $\ell_{3}$ has the largest Yukawa coupling to $N$.
There is an upper bound on $c_{\ell 3}$ from the $\tau$ mass.
The Dirac masses of zero modes are shown in figure \ref{Dirac}.
We see that $c_{\ell 3} \le 0.6$ is necessary to reproduce the $\tau$ mass.

First, we consider Dirac neutrino case.
In this case, $N$ should be localized toward the Planck brane
to obtain a small Yukawa coupling $\sim 10^{-12}$.
From figure \ref{Dirac} we see that
$c_{N} \simeq 1.3$ is required to reproduce the neutrino mass.

Next, we consider Majorana neutrino case.
The mass matrix of the Majorana mass term of $N$ is
\begin{align}
  \begin{bmatrix}
    M P_{0}(c_{N}) P_{0}(c_{N})
  & M P_{0}(c_{N}) P_{1}(c_{N})
  & M P_{0}(c_{N}) P_{2}(c_{N})
  & \cdots \smallskip \\
    M P_{1}(c_{N}) P_{0}(c_{N})
  & M P_{1}(c_{N}) P_{1}(c_{N})
  & \cdots
  & \cdots \\
    M P_{2}(c_{N}) P_{0}(c_{N})
  & \vdots
  & \ddots
  & \cdots \\
    \vdots
  & \vdots
  & \vdots
  & \ddots
  \end{bmatrix},
\end{align}
which is diagonalized to give the eigenvalues
\begin{align}
  \sum_{n=0} M P_{n}(c_{N}) P_{n}(c_{N}), 0, 0, \cdots.
\end{align}
We assume $c_{N} > 1/2$ to ensure that
$M P_{0}(c_{N}) P_{0}(c_{N}) \gg M P_{n}(c_{N}) P_{n}(c_{N})$.
In this case, mixings between the zero mode and the KK modes are small,
and the KK modes are almost Dirac particles.

\begin{figure}[tbp]
\begin{center}
%WinTpicVersion3.08
\unitlength 0.1in
\begin{picture}( 59.2500, 22.0000)(  4.7500,-24.1500)
% LINE 2 2 3 0
% 2 800 800 1200 1400
% 
\special{pn 8}%
\special{pa 800 800}%
\special{pa 1200 1400}%
\special{dt 0.045}%
% LINE 2 0 3 0
% 2 1200 1400 2800 1400
% 
\special{pn 8}%
\special{pa 1200 1400}%
\special{pa 2800 1400}%
\special{fp}%
% LINE 2 2 3 0
% 2 2800 1400 3200 800
% 
\special{pn 8}%
\special{pa 2800 1400}%
\special{pa 3200 800}%
\special{dt 0.045}%
% LINE 2 0 3 0
% 2 1200 1400 800 2000
% 
\special{pn 8}%
\special{pa 1200 1400}%
\special{pa 800 2000}%
\special{fp}%
% LINE 2 0 3 0
% 2 2800 1400 3200 2000
% 
\special{pn 8}%
\special{pa 2800 1400}%
\special{pa 3200 2000}%
\special{fp}%
% LINE 2 0 3 0
% 2 1950 1350 2050 1450
% 
\special{pn 8}%
\special{pa 1950 1350}%
\special{pa 2050 1450}%
\special{fp}%
% LINE 2 0 3 0
% 2 2050 1350 1950 1450
% 
\special{pn 8}%
\special{pa 2050 1350}%
\special{pa 1950 1450}%
\special{fp}%
% STR 2 0 3 0
% 3 700 600 700 700 5 0
% $v$
\put(7.0000,-7.0000){\makebox(0,0){$v$}}%
% STR 2 0 3 0
% 3 3300 600 3300 700 5 0
% $v$
\put(33.0000,-7.0000){\makebox(0,0){$v$}}%
% STR 2 0 3 0
% 3 700 2000 700 2100 5 0
% $\nu$
\put(7.0000,-21.0000){\makebox(0,0){$\nu$}}%
% STR 2 0 3 0
% 3 3300 2000 3300 2100 5 0
% $\nu$
\put(33.0000,-21.0000){\makebox(0,0){$\nu$}}%
% STR 2 0 3 0
% 3 1600 1500 1600 1600 5 0
% $N^{(0)}$
\put(16.0000,-16.0000){\makebox(0,0){$N^{(0)}$}}%
% STR 2 0 3 0
% 3 2400 1500 2400 1600 5 0
% $N^{(0)}$
\put(24.0000,-16.0000){\makebox(0,0){$N^{(0)}$}}%
% STR 2 0 3 0
% 3 2000 1100 2000 1200 5 0
% $M_{00}$
\put(20.0000,-12.0000){\makebox(0,0){$M_{00}$}}%
% LINE 2 0 3 0
% 2 4000 2400 4400 1800
% 
\special{pn 8}%
\special{pa 4000 2400}%
\special{pa 4400 1800}%
\special{fp}%
% LINE 2 0 3 0
% 2 4400 1800 6000 1800
% 
\special{pn 8}%
\special{pa 4400 1800}%
\special{pa 6000 1800}%
\special{fp}%
% LINE 2 0 3 0
% 2 6000 1800 6400 2400
% 
\special{pn 8}%
\special{pa 6000 1800}%
\special{pa 6400 2400}%
\special{fp}%
% CIRCLE 2 2 3 0
% 4 5200 1800 6000 1800 6000 1800 4400 1800
% 
\special{pn 8}%
\special{ar 5200 1800 800 800  3.1415927 3.1565927}%
\special{ar 5200 1800 800 800  3.2015927 3.2165927}%
\special{ar 5200 1800 800 800  3.2615927 3.2765927}%
\special{ar 5200 1800 800 800  3.3215927 3.3365927}%
\special{ar 5200 1800 800 800  3.3815927 3.3965927}%
\special{ar 5200 1800 800 800  3.4415927 3.4565927}%
\special{ar 5200 1800 800 800  3.5015927 3.5165927}%
\special{ar 5200 1800 800 800  3.5615927 3.5765927}%
\special{ar 5200 1800 800 800  3.6215927 3.6365927}%
\special{ar 5200 1800 800 800  3.6815927 3.6965927}%
\special{ar 5200 1800 800 800  3.7415927 3.7565927}%
\special{ar 5200 1800 800 800  3.8015927 3.8165927}%
\special{ar 5200 1800 800 800  3.8615927 3.8765927}%
\special{ar 5200 1800 800 800  3.9215927 3.9365927}%
\special{ar 5200 1800 800 800  3.9815927 3.9965927}%
\special{ar 5200 1800 800 800  4.0415927 4.0565927}%
\special{ar 5200 1800 800 800  4.1015927 4.1165927}%
\special{ar 5200 1800 800 800  4.1615927 4.1765927}%
\special{ar 5200 1800 800 800  4.2215927 4.2365927}%
\special{ar 5200 1800 800 800  4.2815927 4.2965927}%
\special{ar 5200 1800 800 800  4.3415927 4.3565927}%
\special{ar 5200 1800 800 800  4.4015927 4.4165927}%
\special{ar 5200 1800 800 800  4.4615927 4.4765927}%
\special{ar 5200 1800 800 800  4.5215927 4.5365927}%
\special{ar 5200 1800 800 800  4.5815927 4.5965927}%
\special{ar 5200 1800 800 800  4.6415927 4.6565927}%
\special{ar 5200 1800 800 800  4.7015927 4.7165927}%
\special{ar 5200 1800 800 800  4.7615927 4.7765927}%
\special{ar 5200 1800 800 800  4.8215927 4.8365927}%
\special{ar 5200 1800 800 800  4.8815927 4.8965927}%
\special{ar 5200 1800 800 800  4.9415927 4.9565927}%
\special{ar 5200 1800 800 800  5.0015927 5.0165927}%
\special{ar 5200 1800 800 800  5.0615927 5.0765927}%
\special{ar 5200 1800 800 800  5.1215927 5.1365927}%
\special{ar 5200 1800 800 800  5.1815927 5.1965927}%
\special{ar 5200 1800 800 800  5.2415927 5.2565927}%
\special{ar 5200 1800 800 800  5.3015927 5.3165927}%
\special{ar 5200 1800 800 800  5.3615927 5.3765927}%
\special{ar 5200 1800 800 800  5.4215927 5.4365927}%
\special{ar 5200 1800 800 800  5.4815927 5.4965927}%
\special{ar 5200 1800 800 800  5.5415927 5.5565927}%
\special{ar 5200 1800 800 800  5.6015927 5.6165927}%
\special{ar 5200 1800 800 800  5.6615927 5.6765927}%
\special{ar 5200 1800 800 800  5.7215927 5.7365927}%
\special{ar 5200 1800 800 800  5.7815927 5.7965927}%
\special{ar 5200 1800 800 800  5.8415927 5.8565927}%
\special{ar 5200 1800 800 800  5.9015927 5.9165927}%
\special{ar 5200 1800 800 800  5.9615927 5.9765927}%
\special{ar 5200 1800 800 800  6.0215927 6.0365927}%
\special{ar 5200 1800 800 800  6.0815927 6.0965927}%
\special{ar 5200 1800 800 800  6.1415927 6.1565927}%
\special{ar 5200 1800 800 800  6.2015927 6.2165927}%
\special{ar 5200 1800 800 800  6.2615927 6.2765927}%
% LINE 2 2 3 0
% 2 5200 1000 4800 400
% 
\special{pn 8}%
\special{pa 5200 1000}%
\special{pa 4800 400}%
\special{dt 0.045}%
% LINE 2 2 3 0
% 2 5200 1000 5600 400
% 
\special{pn 8}%
\special{pa 5200 1000}%
\special{pa 5600 400}%
\special{dt 0.045}%
% STR 2 0 3 0
% 3 4700 200 4700 300 5 0
% $v$
\put(47.0000,-3.0000){\makebox(0,0){$v$}}%
% STR 2 0 3 0
% 3 5700 200 5700 300 5 0
% $v$
\put(57.0000,-3.0000){\makebox(0,0){$v$}}%
% LINE 2 0 3 0
% 2 5150 1750 5250 1850
% 
\special{pn 8}%
\special{pa 5150 1750}%
\special{pa 5250 1850}%
\special{fp}%
% LINE 2 0 3 0
% 2 5250 1750 5150 1850
% 
\special{pn 8}%
\special{pa 5250 1750}%
\special{pa 5150 1850}%
\special{fp}%
% STR 2 0 3 0
% 3 5200 1500 5200 1600 5 0
% $M_{ij}$
\put(52.0000,-16.0000){\makebox(0,0){$M_{ij}$}}%
% STR 2 0 3 0
% 3 4800 1900 4800 2000 5 0
% $N^{(i)}$
\put(48.0000,-20.0000){\makebox(0,0){$N^{(i)}$}}%
% STR 2 0 3 0
% 3 5600 1900 5600 2000 5 0
% $N^{(j)}$
\put(56.0000,-20.0000){\makebox(0,0){$N^{(j)}$}}%
% STR 2 0 3 0
% 3 4400 1100 4400 1200 5 0
% $H^{0}$
\put(44.0000,-12.0000){\makebox(0,0){$H^{0}$}}%
% STR 2 0 3 0
% 3 6000 1100 6000 1200 5 0
% $H^{0}$
\put(60.0000,-12.0000){\makebox(0,0){$H^{0}$}}%
% STR 2 0 3 0
% 3 3900 2400 3900 2500 5 0
% $\nu$
\put(39.0000,-25.0000){\makebox(0,0){$\nu$}}%
% STR 2 0 3 0
% 3 6500 2400 6500 2500 5 0
% $\nu$
\put(65.0000,-25.0000){\makebox(0,0){$\nu$}}%
\end{picture}%
\caption{
  The diagrams that contribute to the effective neutrino mass.
  Here we use the notation $M_{ij} = \lambda' P_{i}(c_{N}) P_{j}(c_{N}) M$.
}
\label{effdiag}
\end{center}
\end{figure}
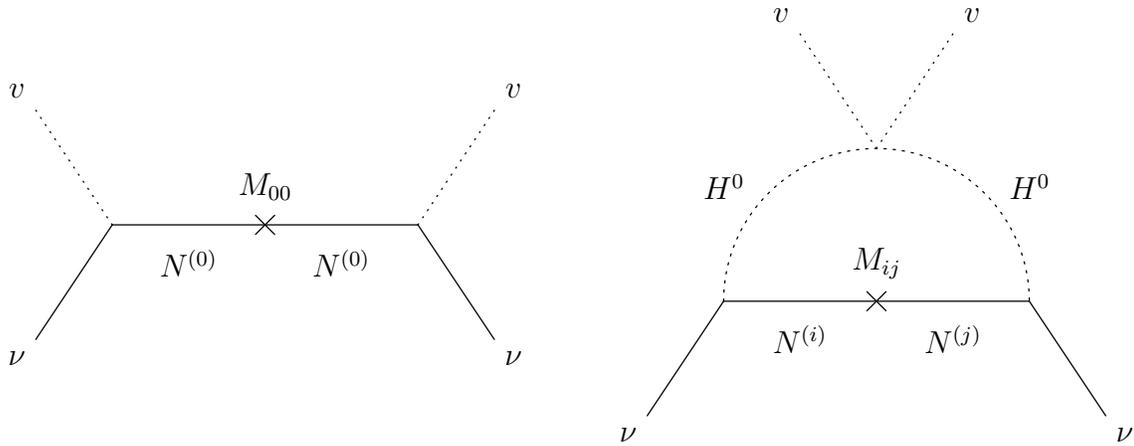

The effective left-handed neutrino mass is generated
by both the seesaw mechanism and 1-loop effects,
which are shown in figure \ref{effdiag}.
The KK modes do not contribute to the seesaw mechanism, since they are Dirac particles.
The effective mass obtained from the seesaw mechanism is
\begin{align}
  m_\mathrm{seesaw}
= \frac{[\lambda_{\nu 5} T_{0}(c_{\ell}) T_{0}(c_{N}) \tilde{H}]^{2}}
       {\lambda' P_{0}(c_{N})^{2} M}.
  \label{seesaw}
\end{align}
The contribution from 1-loop effects is
\begin{align}
  m_\mathrm{loop}
  \sim
  \frac{1}{16 \pi^{2}}
  \frac{\pi^{2}}{12}
  \frac{[\lambda_{\nu 5} T_{0}(c_{\ell}) T_{1}(c_{N}) \tilde{H}]^{2}
        [\lambda' P_{1}(c_{N})^{2} M]}
       {m_{1}^{2}},
  \label{loop}
\end{align}
where we have summed over all the KK modes under the approximations
\begin{align}
P_{n}(c_{N}) & = -(-1)^{n}P_{1}(c_{N}), \\
T_{n}(c_{N}) & = T_{1}(c_{N}), \\
m_{N}^{(n)}  & = n m_{N}^{(1)},
\end{align}
which are confirmed numerically.

\begin{figure}[tbp]
\begin{center}
\includegraphics[width = 0.45 \linewidth]{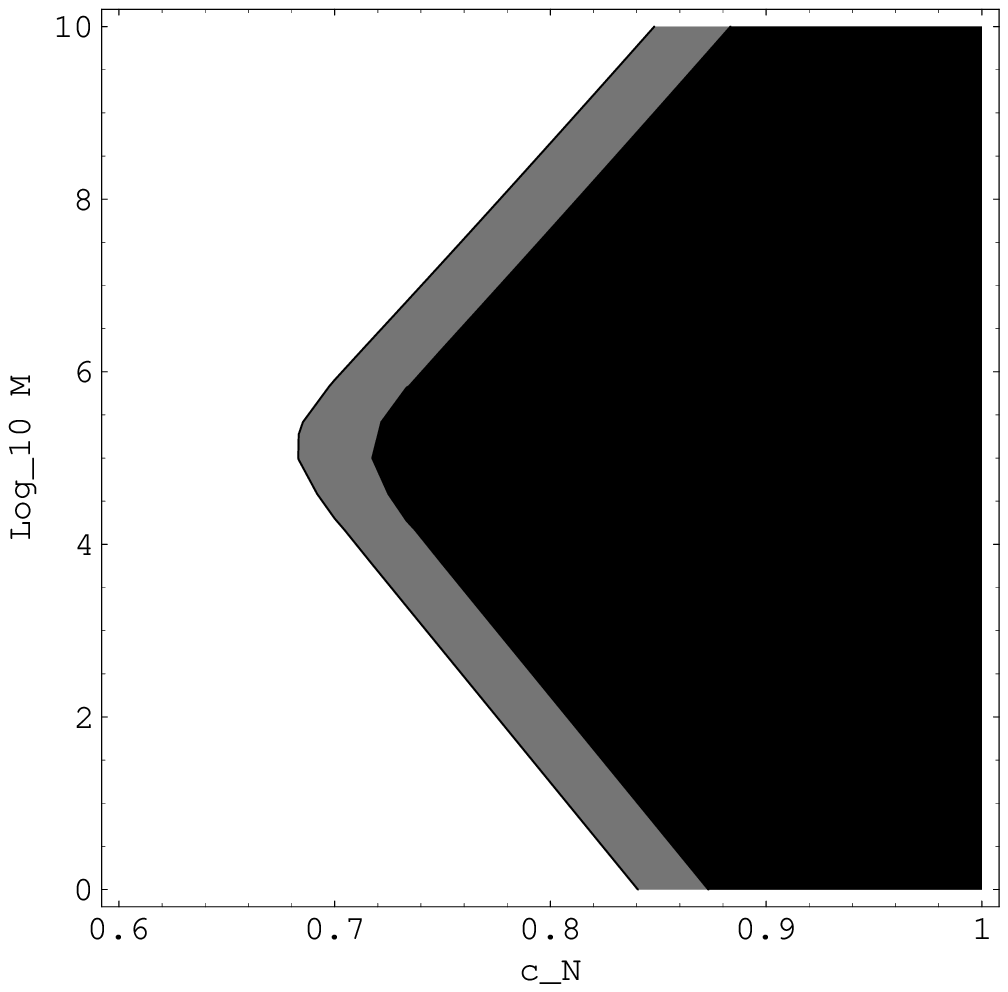}
\includegraphics[width = 0.45 \linewidth]{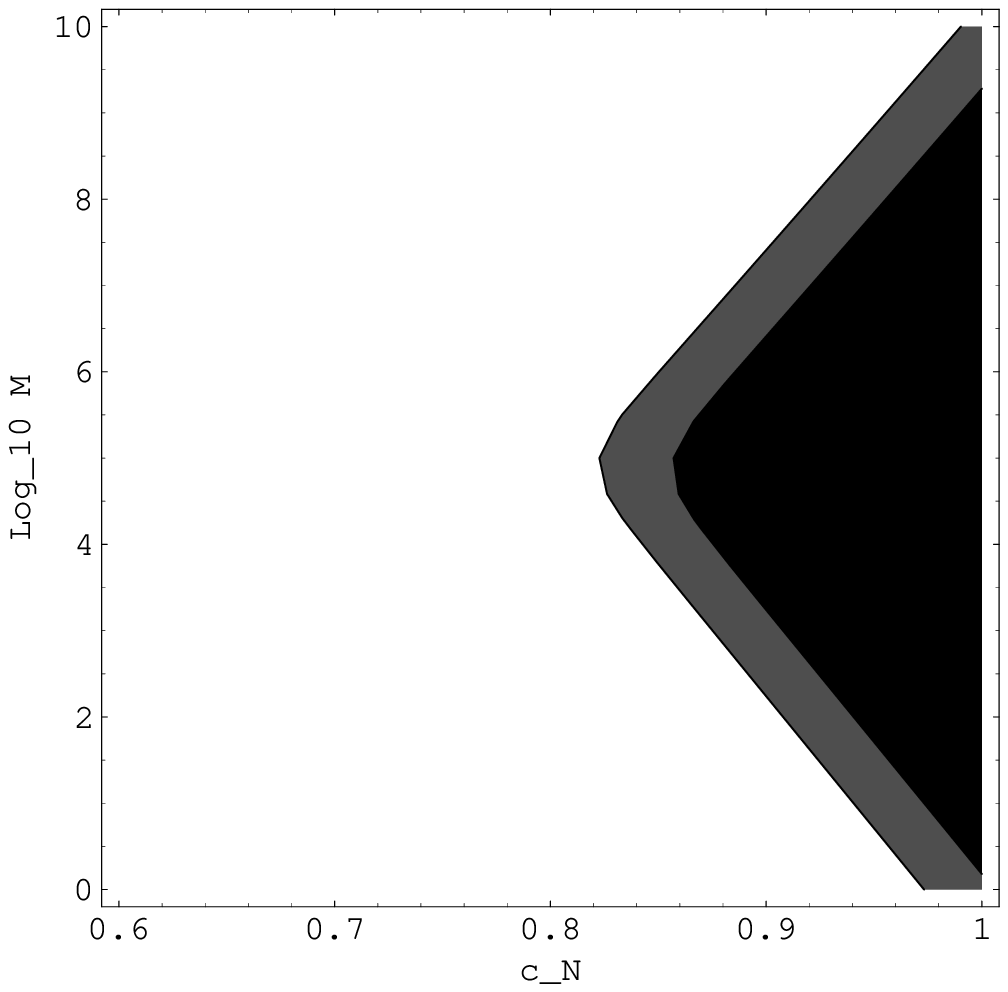}
\caption{
  The effective neutrino mass.
  The figure on the left is for $c_{\ell 3} = 0.6$
 (upper bound from $\tau$ mass),
  and that on the right for $c_{\ell 3} = -0.5$.
  The two contours correspond to $10^{-1}$ eV (left) and $10^{-2}$ eV (right).
}
\label{effnumass}
\end{center}
\end{figure}

The sum of these two contributions is plotted in figure \ref{effnumass}.
The effective neutrino mass is determined
by the seesaw mechanism for $M < 10^{5}$ GeV,
and by 1-loop effects for $M > 10^{5}$ GeV.
The parameter $c_{N}$ can be as low as 0.7,
which is in contrast with $c_{N} \simeq 1.3$ for Dirac neutrino case.

Now we discuss the relation between the mass spectrum of the KK modes
and configurations of fermions.
The mass of the 1st KK mode of fermion is
\begin{align}
  m_{\psi}^{(1)} \simeq 2.4 k e^{- k \pi R} \, [1 + 0.6|c - 1/2|],
\end{align}
and the mass for $c = 1/2$ is almost equal to the KK gauge boson mass $m_{A}^{(1)}$.
Thus we can evaluate the $c$ parameter from the ratio $m_{\psi}^{(1)} / m_{A}^{(1)}$.
We showed that for Dirac neutrino case $c_{N} \simeq 1.3$,
while for Majorana neutrino case $c_{N} \ge 0.7$.
Hence we can see whether neutrino masses are Dirac type or Majorana type
from measurements of the KK masses $m_{A}^{(1)}$ and $m_{N}^{(1)}$.

%%%%%%%%%%%%%%%%%%%%%%%%%%%%%%%%%%%%%%%%%%%%%%%%%%

\section{Collider signal}

In 4D theories, the production of right-handed neutrinos at colliders is difficult,
since their masses are large (for Majorana case),
or their Yukawa couplings are small (for both Dirac and Majorana cases).
In the RS model, the 1st KK modes of fermions have masses of order TeV,
and Yukawa couplings of order 1.
Thus the 1st KK mode of right-handed neutrino $N^{(1)}$
may be produced at TeV colliders,
whether the neutino mass is Dirac or Majorana type.
In the following, we discuss how to produce $N^{(1)}$
for two cases, $c_{\ell 1} < 1/2$ and $c_{\ell 1} > 1/2$.

%%%%%%%%%%%%%%%%%%%%%%%%%

\subsection{$c_{\ell 1} < 1/2$}

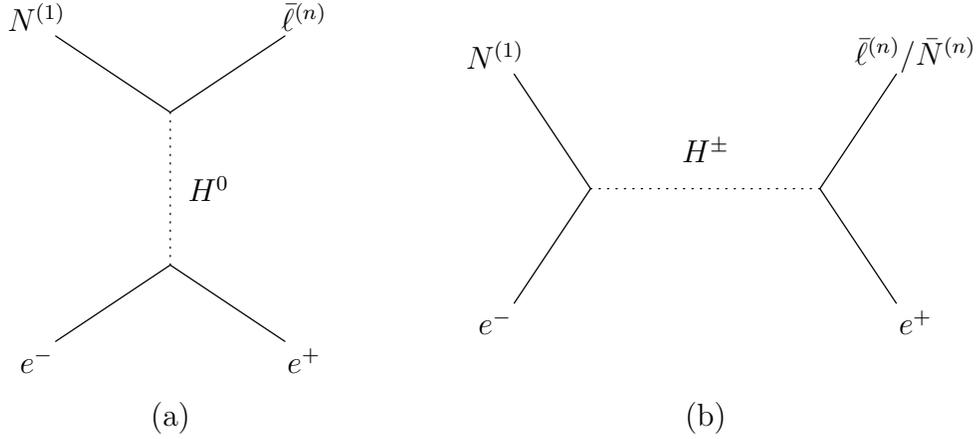
\begin{figure}[tbp]
\begin{center}
%WinTpicVersion3.08
\unitlength 0.1in
\begin{picture}( 49.0500, 21.0000)(  2.9500,-25.1500)
% LINE 2 0 3 0
% 2 800 2200 1400 1800
% 
\special{pn 8}%
\special{pa 800 2200}%
\special{pa 1400 1800}%
\special{fp}%
% LINE 2 0 3 0
% 2 1400 1800 2000 2200
% 
\special{pn 8}%
\special{pa 1400 1800}%
\special{pa 2000 2200}%
\special{fp}%
% LINE 2 2 3 0
% 2 1400 1800 1400 1000
% 
\special{pn 8}%
\special{pa 1400 1800}%
\special{pa 1400 1000}%
\special{dt 0.045}%
% LINE 2 0 3 0
% 2 1400 1000 800 600
% 
\special{pn 8}%
\special{pa 1400 1000}%
\special{pa 800 600}%
\special{fp}%
% LINE 2 0 3 0
% 2 1400 1000 2000 600
% 
\special{pn 8}%
\special{pa 1400 1000}%
\special{pa 2000 600}%
\special{fp}%
% STR 2 0 3 0
% 3 700 400 700 500 5 0
% $N^{(1)}$
\put(7.0000,-5.0000){\makebox(0,0){$N^{(1)}$}}%
% STR 2 0 3 0
% 3 700 2200 700 2300 5 0
% $e^{-}$
\put(7.0000,-23.0000){\makebox(0,0){$e^{-}$}}%
% STR 2 0 3 0
% 3 2100 2200 2100 2300 5 0
% $e^{+}$
\put(21.0000,-23.0000){\makebox(0,0){$e^{+}$}}%
% LINE 2 0 3 0
% 2 3200 800 3600 1400
% 
\special{pn 8}%
\special{pa 3200 800}%
\special{pa 3600 1400}%
\special{fp}%
% LINE 2 0 3 0
% 2 3600 1400 3200 2000
% 
\special{pn 8}%
\special{pa 3600 1400}%
\special{pa 3200 2000}%
\special{fp}%
% LINE 2 2 3 0
% 2 3600 1400 4800 1400
% 
\special{pn 8}%
\special{pa 3600 1400}%
\special{pa 4800 1400}%
\special{dt 0.045}%
% LINE 2 0 3 0
% 2 4800 1400 5200 800
% 
\special{pn 8}%
\special{pa 4800 1400}%
\special{pa 5200 800}%
\special{fp}%
% LINE 2 0 3 0
% 2 4800 1400 5200 2000
% 
\special{pn 8}%
\special{pa 4800 1400}%
\special{pa 5200 2000}%
\special{fp}%
% STR 2 0 3 0
% 3 3100 2000 3100 2100 5 0
% $e^{-}$
\put(31.0000,-21.0000){\makebox(0,0){$e^{-}$}}%
% STR 2 0 3 0
% 3 5300 2000 5300 2100 5 0
% $e^{+}$
\put(53.0000,-21.0000){\makebox(0,0){$e^{+}$}}%
% STR 2 0 3 0
% 3 3100 600 3100 700 5 0
% $N^{(1)}$
\put(31.0000,-7.0000){\makebox(0,0){$N^{(1)}$}}%
% STR 2 0 3 0
% 3 5300 600 5300 700 5 0
% $\bar{\ell}^{(n)}/\bar{N}^{(n)}$
\put(53.0000,-7.0000){\makebox(0,0){$\bar{\ell}^{(n)}/\bar{N}^{(n)}$}}%
% STR 2 0 3 0
% 3 1600 1300 1600 1400 5 0
% $H^{0}$
\put(16.0000,-14.0000){\makebox(0,0){$H^{0}$}}%
% STR 2 0 3 0
% 3 4200 1100 4200 1200 5 0
% $H^{\pm}$
\put(42.0000,-12.0000){\makebox(0,0){$H^{\pm}$}}%
% STR 2 0 3 0
% 3 1400 2500 1400 2600 5 0
% (a)
\put(14.0000,-26.0000){\makebox(0,0){(a)}}%
% STR 2 0 3 0
% 3 4200 2500 4200 2600 5 0
% (b)
\put(42.0000,-26.0000){\makebox(0,0){(b)}}%
% STR 2 0 3 0
% 3 2100 400 2100 500 5 0
% $\bar{\ell}^{(n)}$
\put(21.0000,-5.0000){\makebox(0,0){$\bar{\ell}^{(n)}$}}%
\end{picture}%
\caption{
  The diagrams of processes including $N^{(1)}$ in the final state.
}
\label{ee2NN}
\end{center}
\end{figure}

\begin{figure}[tbp]
\begin{center}
\includegraphics{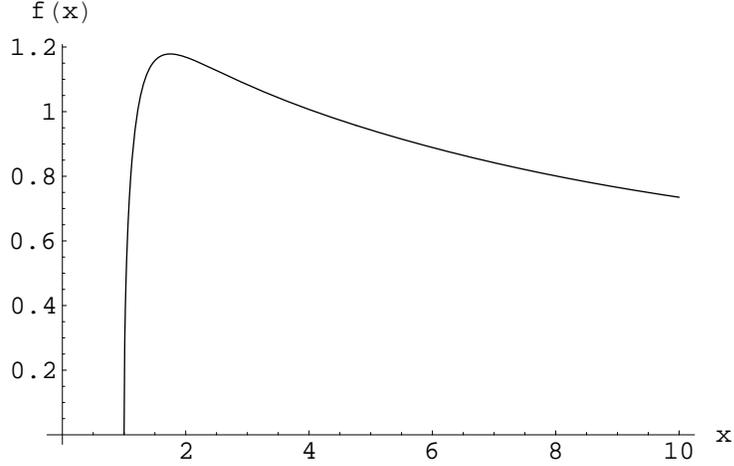}
\caption{
  The function $f(x)$ of (\ref{sigma}).
}
\label{Fx}
\end{center}
\end{figure}

In this case, the Yukawa coupling between $N^{(1)}$ and electron is
\begin{align}
  \lambda
= \lambda_{\nu 5} T_{0}(c_{\ell 1}) T_{1}(c_{N})
  \simeq \sqrt{1/2 - c_{\ell 1}}
  \sim 1.
\end{align}
Thus $N^{(1)}$ can be produced by the Yukawa interactions.
The diagrams of corresponding processes are shown in figure \ref{ee2NN}.
The reaction $e^{-} e^{+} \to N^{(1)} \bar{N}^{(1)}$ has the largest cross section
\begin{align}
  \sigma(e^{-} e^{+} \to N^{(1)} \bar{N}^{(1)})
= 110 \ \mathrm{fb} \times
  \left( \frac{\lambda}{1.0} \right)^{4}
  \left( \frac{6 \ \mathrm{TeV}}{E_\mathrm{cm}} \right)^{2}
  f(E_\mathrm{cm}/2m_{N}^{(1)}),
  \label{sigma}
\end{align}
where $f(x)$ is shown in figure \ref{Fx}.
The reaction $e^{-} e^{+} \to N^{(1)} \bar{N}^{(0)}$ is negligible,
since it is suppressed by the Yukawa coupling between $N^{(0)}$ and electron
\begin{align}
       \lambda_{\nu 5} T_{0}(c_{\ell 1}) T_{0}(c_{N})
\simeq \sqrt{1/2 - c_{\ell 1}} \sqrt{c_{N} - 1/2} \ e^{(1/2 - c_{N}) k \pi R}
   \ll 1,
\end{align}
for both Dirac and Majorana neutrino cases.
The processes (a) are suppressed
by the electron Yukawa coupling $m_{e}/v \sim 10^{-6}$.
The processes (b) with final states $N^{1} \bar{\ell}^{(n)}$
are also suppressed by $m_{e}/v$,
since the chiralities of initial electrons are $e_{L}^{-}$ and $e_{R}^{+}$.

The decay modes of $N^{(1)}$ depend on $c_{N}$, $c_{\ell i}$ and $c_{ei}$,
which determine the couplings and the KK masses of leptons.
We consider the simplest case $m_{N}^{(1)} < m_{\ell i}^{(1)}$.
The only decay mode of $N^{(1)}$ is
\begin{align}
  N^{(1)} \to \ell_{i}^{(0)} + H.
\end{align}
Thus we observe a lepton and a Higgs both with energy $\sim E_\mathrm{cm}/4$,
which is specific to the decay of $N^{(1)}$.

%%%%%%%%%%%%%%%%%%%%%%%%%

\subsection{$c_{\ell 1} > 1/2$}

\begin{figure}[tbp]
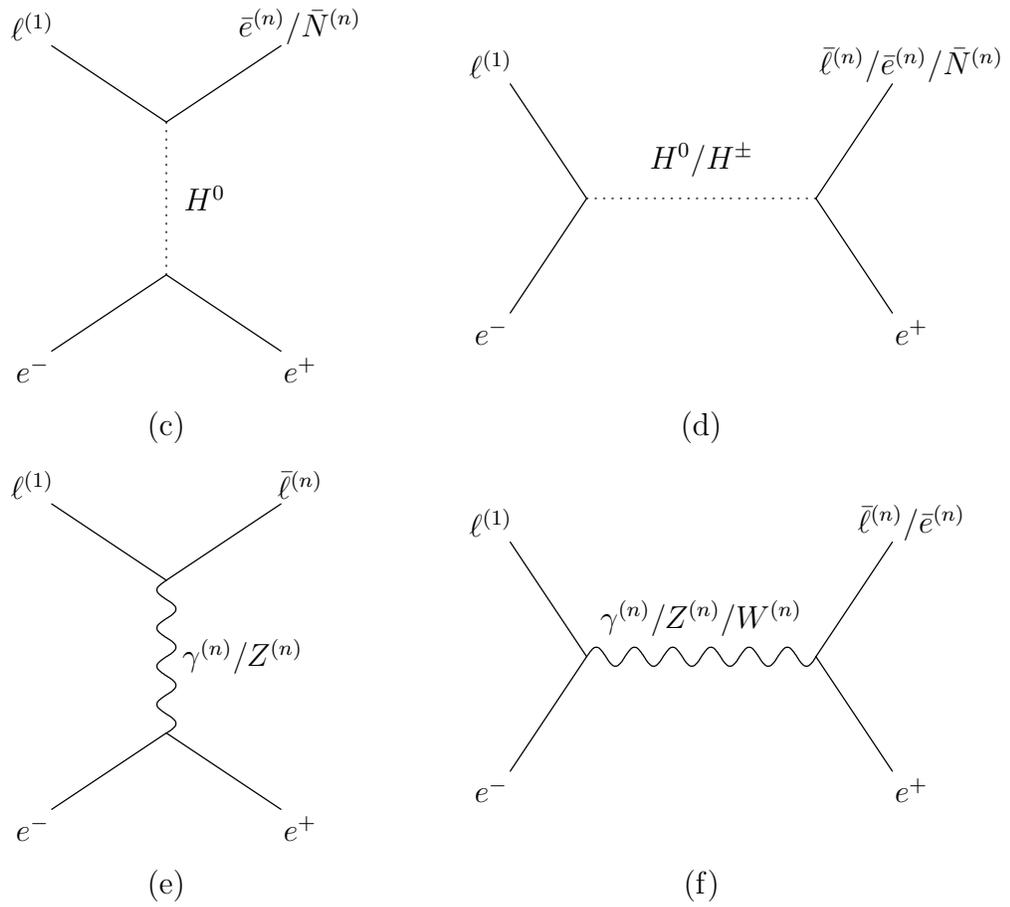

\begin{center}
\include{ee2LL}
\caption{
  The diagrams of processes including $\ell^{(1)}$ in the final state.
}
\label{ee2LL}
\end{center}
\end{figure}

In this case, the Yukawa coupling between $N^{(1)}$ and electron is
\begin{align}
  \lambda_{5} T_{0}(c_{\ell 1}) T_{1}(c_{N})
  \simeq \sqrt{c_{\ell 1} - 1/2} \ e^{(1/2 - c_{\ell 1}) k \pi R}
  \ll 1.
\end{align}
Thus we can not produce $N^{(1)}$ by the Yukawa interactions.
However, the 1st KK modes of lepton doublets $\ell_{i}^{(1)}$
are produced by the processes of figure \ref{ee2LL}.
Assuming the mass ordering $m_{N}^{(1)} < m_{\ell i}^{(1)} < m_{e i}^{(1)}$,
the main decay mode of $\ell_{i}^{(1)}$ is
\begin{align}
 \ell_{i}^{(1)} \to N^{(1)} + H,
\end{align}
since the Yukawa couplings between $\ell_{i}^{(1)}$ and $N^{(1)}$ are
$\lambda_{\nu 5} \sqrt{k} \sqrt{k} \sim 1$.

For the case $c_{e1} < 1/2$,
the Yukawa couplings between $\ell_{1}^{(1)}$ and electron are order 1.
Thus the processes
\begin{align}
  e^{-} e^{+} & \to \ell_{1}^{(1)} \bar{\ell}_{1}^{(n)} \qquad (n \ge 1)
  \label{ee2ll}
\end{align}
occur with considerable rate, while other processes are suppressed
by gauge couplings or the electron Yukawa coupling.
The processes (\ref{ee2ll}) with $n \ge 2$ are kinematically suppressed.
The cross section of
$e^{-} e^{+} \to \ell_{1}^{(1)} \bar{\ell}_{1}^{(1)}$
is almost equal to (\ref{sigma}).

For the case $c_{e1} > 1/2$,
the Yukawa couplings between the KK modes of leptons and electron are much smaller than 1.
Thus $\ell_{i}^{(1)}$ are produced by electroweak interactions (processes (e) and (f))
with cross sections of $1 - 10$ fb for $E_\mathrm{cm} \sim 6$ TeV,
depending on the configurations and the KK masses of fermions.

%%%%%%%%%%%%%%%%%%%%%%%%%%%%%%%%%%%%%%%%%%%%%%%%%%

\section{Conclusion}

We have shown in section 4 that the 1st KK mode of right-handed neutrino $N^{(1)}$
can be produced at $e^{+} e^{-}$ colliders with $E_\mathrm{cm} > 6$ TeV.
In particular, if the left-handed electron $\ell_{1}^{(0)}$
is localized toward the TeV brane,
then $N^{(1)}$ is produced with a considerable rate.
From the measurement of masses of KK gauge bosons and $N^{(1)}$,
we can see wheather neutrinos are Dirac or Majorana particles,
as we discussed in section 3.
This is a remarkable feature in the RS model,
since in 4D theories it is difficult to distinguish the type of neutrino mass,
which is related to the origin of matter in the universe \cite{Fukugita:1986hr}.

%%%%%%%%%%%%%%%%%%%%%%%%%%%%%%%%%%%%%%%%%%%%%%%%%%

\section*{Acknowledgement}

We thank T. T. Yanagida for stimulating discussions.
We also thank K. Hamaguchi for careful reading of the manuscript.

%%%%%%%%%%%%%%%%%%%%%%%%%%%%%%%%%%%%%%%%%%%%%%%%%%

\appendix
\section*{Appendix}

We work in the gauge where
\begin{align}
  0 & = A_{5}(x,y), \\
  0 & = \eta^{\mu \nu} \partial_{\mu} A_{\nu}(x,y).
\end{align}
In this gauge, the zero mode transforms as a gauge field,
while the KK modes transform as adjoint fields.
The gauge covariant derivative is
\begin{align}
    D_{M}
& = \partial_{M} + i g_{5} A_{M}(x,y) \notag \\
& = \partial_{M} + i g_{4} A_{\mu}^{(0)}(x) + \textrm{(KK modes)},
\end{align}
where $g_{5} = g_{4} \sqrt{2 \pi R}$ \cite{Davoudiasl:1999tf, Pomarol:1999ad}.

The 5D fermions are decomposed to the KK modes \cite{Grossman:1999ra},
\begin{align}
  \Psi(x,y)
= \begin{bmatrix}
    \psi_{L}^{(n)}(x) f_{L}^{(n)}(y) \\
    \psi_{R}^{(n)}(x) f_{R}^{(n)}(y)
  \end{bmatrix}
= e^{(3/2) k y}
  \begin{bmatrix}
    \psi_{L}^{(n)}(x) \hat{f}_{L}^{(n)}(y) \\
    \psi_{R}^{(n)}(x) \hat{f}_{R}^{(n)}(y)
  \end{bmatrix},
\end{align}
and $\hat{f}_{L}^{(n)}(y)$, $\hat{f}_{R}^{(n)}(y)$
satisfies the normalization condition,
\begin{align}
  \delta_{mn}
= \int_{- \pi R}^{\pi R} dy \, \hat{f}_{L}^{(m)}(y) \hat{f}_{L}^{(n)}(y)
= \int_{- \pi R}^{\pi R} dy \, \hat{f}_{R}^{(m)}(y) \hat{f}_{R}^{(n)}(y).
\end{align}
The $Z_{2}$ transformation is given by $\Psi(x,-y) = \pm \gamma_{5} \Psi(x,y)$.
This property ensures that the zero modes are chiral.
The KK modes form Dirac fields,
\begin{align}
  \psi^{(n)}(x)
= \begin{bmatrix}
    \psi_{L}^{(n)}(x) \\
    \psi_{R}^{(n)}(x)
  \end{bmatrix}.
\end{align}
We denote the profile of the $n$th even component as $f^{(n)}(y)$:
$f^{(n)}(y) = f_{L}^{(n)}(y)$ or $f_{R}^{(n)}(y)$,
depending on whether the zero mode is left- or right-handed.
We introduce functions $P_{n}(c)$ and $T_{n}(c)$ defined by
\begin{align}
  P_{n}(c) & \equiv \hat{f}^{(n)}(0), \\
  T_{n}(c) & \equiv \hat{f}^{(n)}(\pi R).
\end{align}
The 5D configuration depends on the bulk mass $m_\mathrm{bulk}$.
The configuration of the zero mode is
\begin{align}
    \hat{f}^{(0)}(y)
& = \sqrt{\dfrac{(1/2 - c) k}{e^{(1 - 2c) k \pi R} - 1}} \ e^{(1/2 - c) k y} \\
& \simeq
  \left\{
  \begin{array}{rll}
    \sqrt{(1/2 - c) k} \! & \! e^{(1/2 - c) k (y - \pi R)}
  & \mathrm{for} \ c < 1/2, \smallskip \\
    \sqrt{1 / 2 \pi R} \! & \! e^{(1/2 - c) k y}
  & \mathrm{for} \ c \simeq 1/2, \smallskip \\
    \sqrt{(c - 1/2) k} \! & \! e^{(1/2 - c) k y}
  & \mathrm{for} \ c > 1/2,
  \end{array}
  \right.
  \label{fzero}
\end{align}
where $c = m_\mathrm{bulk}/k$.
The couplings of the 1st KK mode on the branes are
\begin{align}
  T_{1}(c)
& \simeq \sqrt{k}, \\
  P_{1}(c)
& \simeq
  \left\{
  \begin{array}{ll}
    (1.4 - 2c) P_{0}(c) & \textrm{for} \ c < 1/2, \smallskip \\
    (0.6 - 2c) T_{0}(c) & \textrm{for} \ c > 1/2.
  \end{array}
  \right.
  \label{P1}
\end{align}

%%%%%%%%%%%%%%%%%%%%%%%%%%%%%%%%%%%%%%%%%%%%%%%%%%


\begin{thebibliography}{99}

%\cite{Randall:1999ee}
\bibitem{Randall:1999ee}
  L.~Randall and R.~Sundrum,
  %``A large mass hierarchy from a small extra dimension,''
  Phys.\ Rev.\ Lett.\  {\bf 83}, 3370 (1999)
  %[arXiv:hep-ph/9905221].
  %%CITATION = PRLTA,83,3370;%%

%%%%%%%%%%%%%%%%%%%%%%%%%
%%%%%  Unification  %%%%%
%%%%%%%%%%%%%%%%%%%%%%%%%

%\cite{Pomarol:2000hp}
\bibitem{Pomarol:2000hp}
  A.~Pomarol,
  %``Grand unified theories without the desert,''
  Phys.\ Rev.\ Lett.\  {\bf 85}, 4004 (2000)
  %[arXiv:hep-ph/0005293].
  %%CITATION = PRLTA,85,4004;%%

%\cite{Randall:2001gb}
\bibitem{Randall:2001gb}
  L.~Randall and M.~D.~Schwartz,
  %``Quantum field theory and unification in AdS5,''
  JHEP {\bf 0111}, 003 (2001)
  %[arXiv:hep-th/0108114].
  %%CITATION = JHEPA,0111,003;%%

%\cite{Randall:2001gc}
\bibitem{Randall:2001gc}
  L.~Randall and M.~D.~Schwartz,
  %``Unification and the hierarchy from AdS5,''
  Phys.\ Rev.\ Lett.\  {\bf 88}, 081801 (2002)
  %[arXiv:hep-th/0108115].
  %%CITATION = PRLTA,88,081801;%%

%\cite{Goldberger:2002hb}
\bibitem{Goldberger:2002hb}
  W.~D.~Goldberger and I.~Z.~Rothstein,
  %``Effective field theory and unification in AdS backgrounds,''
  Phys.\ Rev.\  D {\bf 68}, 125011 (2003)
  %[arXiv:hep-th/0208060].
  %%CITATION = PHRVA,D68,125011;%%

%\cite{Choi:2002ps}
\bibitem{Choi:2002ps}
  K.~w.~Choi and I.~W.~Kim,
  %``One loop gauge couplings in AdS(5),''
  Phys.\ Rev.\  D {\bf 67}, 045005 (2003)
  %[arXiv:hep-th/0208071].
  %%CITATION = PHRVA,D67,045005;%%

%\cite{Goldberger:2002pc}
\bibitem{Goldberger:2002pc}
  W.~D.~Goldberger, Y.~Nomura and D.~R.~Smith,
  %``Warped supersymmetric grand unification,''
  Phys.\ Rev.\  D {\bf 67}, 075021 (2003)
  %[arXiv:hep-ph/0209158].
  %%CITATION = PHRVA,D67,075021;%%

%\cite{Agashe:2002pr}
\bibitem{Agashe:2002pr}
  K.~Agashe, A.~Delgado and R.~Sundrum,
  %``Grand unification in RS1,''
  Annals Phys.\  {\bf 304}, 145 (2003)
  %[arXiv:hep-ph/0212028].
  %%CITATION = APNYA,304,145;%%

%%%%%%%%%%%%%%%%%%%%%%%%%
%%%%%  Dark Matter  %%%%%
%%%%%%%%%%%%%%%%%%%%%%%%%

%\cite{Agashe:2004ci}
\bibitem{Agashe:2004ci}
  K.~Agashe and G.~Servant,
  %``Warped unification, proton stability and dark matter,''
  Phys.\ Rev.\ Lett.\  {\bf 93}, 231805 (2004)
  %[arXiv:hep-ph/0403143].
  %%CITATION = PRLTA,93,231805;%%

%\cite{Agashe:2004bm}
\bibitem{Agashe:2004bm}
  K.~Agashe and G.~Servant,
  %``Baryon number in warped GUTs: Model building and (dark matter related)
  %phenomenology,''
  JCAP {\bf 0502}, 002 (2005)
  %[arXiv:hep-ph/0411254].
  %%CITATION = JCAPA,0502,002;%%

%%%%%%%%%%%%%%%%%%%%%%%%%
%%%%%  Mass Matrix  %%%%%
%%%%%%%%%%%%%%%%%%%%%%%%%

%\cite{Gherghetta:2000qt}
\bibitem{Gherghetta:2000qt}
  T.~Gherghetta and A.~Pomarol,
  %``Bulk fields and supersymmetry in a slice of AdS,''
  Nucl.\ Phys.\  B {\bf 586}, 141 (2000)
  %[arXiv:hep-ph/0003129].
  %%CITATION = NUPHA,B586,141;%%

%\cite{Dooling:2000ky}
\bibitem{Dooling:2000ky}
  D.~Dooling and K.~Kang,
  %``Gravitational origin of quark masses in an extra-dimensional brane
  %world,''
  Phys.\ Lett.\  B {\bf 502}, 189 (2001)
  %[arXiv:hep-ph/0009307].
  %%CITATION = PHLTA,B502,189;%%

%%%%%%%%%%%%%%%%%%%%%%%%%
%%%%%  Other works  %%%%%
%%%%%%%%%%%%%%%%%%%%%%%%%

%\cite{Moreau:2006np}
\bibitem{Moreau:2006np}
  G.~Moreau and J.~I.~Silva-Marcos,
  %``Flavour physics of the RS model with KK masses reachable at LHC,''
  JHEP {\bf 0603}, 090 (2006)
  %[arXiv:hep-ph/0602155].
  %%CITATION = JHEPA,0603,090;%%

%\cite{Nakajima:2007uk}
\bibitem{Nakajima:2007uk}
  H.~Nakajima and Y.~Shinbara,
  %``Solutions to large B and L breaking in the Randall-Sundrum model,''
  Phys.\ Lett.\  B {\bf 648}, 294 (2007)
  %[arXiv:hep-ph/0702063].
  %%CITATION = PHLTA,B648,294;%%

%\cite{Ibanez:1991hv}
\bibitem{Ibanez:1991hv}
  L.~E.~Ibanez and G.~G.~Ross,
  %``Discrete gauge symmetry anomalies,''
  Phys.\ Lett.\  B {\bf 260}, 291 (1991).
  %%CITATION = PHLTA,B260,291;%%

%\cite{seesaw}
\bibitem{seesaw}
  T. Yanagida,
  in \textit{Proc. Workshop on the Unified Theory and Baryon Number in the Universe},
  ed. by O. Sawada, A. Sugamoto
  (KEK report 79-18, 1979), p. 95;
  M. Gell-Mann, P. Ramond, R. Slansky,
  in \textit{Supergravity},
  ed. by P. van Nieuwenhuizen, D.Z. Freedman
  (North Holland, Amsterdam 1979), p. 315.
  See also
  P.~Minkowski,
  Phys.\ Lett.\  B {\bf 67}, 421 (1977).

%\cite{Agashe:2003zs}
\bibitem{Agashe:2003zs}
  K.~Agashe, A.~Delgado, M.~J.~May and R.~Sundrum,
  %``RS1, custodial isospin and precision tests,''
  JHEP {\bf 0308}, 050 (2003)
  %[arXiv:hep-ph/0308036].
  %%CITATION = JHEPA,0308,050;%%

%\cite{Fukugita:1986hr}
\bibitem{Fukugita:1986hr}
  M.~Fukugita and T.~Yanagida,
  %``Baryogenesis Without Grand Unification,''
  Phys.\ Lett.\  B {\bf 174}, 45 (1986).
  %%CITATION = PHLTA,B174,45;%%

%\cite{Davoudiasl:1999tf}
\bibitem{Davoudiasl:1999tf}
  H.~Davoudiasl, J.~L.~Hewett and T.~G.~Rizzo,
  %``Bulk gauge fields in the Randall-Sundrum model,''
  Phys.\ Lett.\  B {\bf 473}, 43 (2000)
  %[arXiv:hep-ph/9911262].
  %%CITATION = PHLTA,B473,43;%%

%\cite{Pomarol:1999ad}
\bibitem{Pomarol:1999ad}
  A.~Pomarol,
  %``Gauge bosons in a five-dimensional theory with localized gravity,''
  Phys.\ Lett.\  B {\bf 486}, 153 (2000)
  %[arXiv:hep-ph/9911294].
  %%CITATION = PHLTA,B486,153;%%

%\cite{Grossman:1999ra}
\bibitem{Grossman:1999ra}
  Y.~Grossman and M.~Neubert,
  %``Neutrino masses and mixings in non-factorizable geometry,''
  Phys.\ Lett.\  B {\bf 474}, 361 (2000)
  %[arXiv:hep-ph/9912408].
  %%CITATION = PHLTA,B474,361;%%

\end{thebibliography}
\end{document}